\documentclass[conference]{IEEEtran}
\IEEEoverridecommandlockouts
\usepackage{cite}
\usepackage{amsmath,amssymb,amsfonts}
\usepackage{algorithmic}
\usepackage{comment}
\usepackage{graphicx}
\usepackage{multirow}
\usepackage{textcomp}
\usepackage{colortbl}
\usepackage{xcolor}
\usepackage{array}          
\usepackage{booktabs}       
\usepackage{makecell}
\usepackage{amssymb}
\makeatletter

\newcommand{\Rmnum}[1]{\expandafter\@slowromancap\romannumeral #1@}
\makeatother
\def\BibTeX{{\rm B\kern-.05em{\sc i\kern-.025em b}\kern-.08em
    T\kern-.1667em\lower.7ex\hbox{E}\kern-.125emX}}
\DeclareRobustCommand*{\IEEEauthorrefmark}[1]{%
    \raisebox{0pt}[0pt][0pt]{\textsuperscript{\footnotesize\ensuremath{#1}}}}
\begin{document}

\setlength{\floatsep}{5pt plus 2pt minus 2pt}
\setlength{\textfloatsep}{5pt plus 2pt minus 2pt}
\setlength{\intextsep}{5pt plus 2pt minus 2pt}

\title{AnalogTester: A Large Language Model-Based Framework for Automatic Testbench Generation in Analog Circuit Design \\
}

\author{
	\IEEEauthorblockN{
		Weiyu Chen\IEEEauthorrefmark{1,2}, 
		Chengjie Liu\IEEEauthorrefmark{1,2}, 
		Wenhao Huang\IEEEauthorrefmark{3}, 
		Jinyang Lyu\IEEEauthorrefmark{3},\\
        Mingqian Yang\IEEEauthorrefmark{4},
		Yuan Du\IEEEauthorrefmark{1},
        Li Du\IEEEauthorrefmark{*1},
        and Jun Yang\IEEEauthorrefmark{2,5}}
	\IEEEauthorblockA{\IEEEauthorrefmark{1}School of Electronic Science and Engineering, Nanjing University, Nanjing, China}
    \IEEEauthorblockA{\IEEEauthorrefmark{2}National Center of Technology Innovation for EDA, Nanjing, China}
	\IEEEauthorblockA{\IEEEauthorrefmark{3}School of Integrated Circuits, Nanjing University, Nanjing, China}
	\IEEEauthorblockA{\IEEEauthorrefmark{4}Beijing Microelectronics Technology Institute, Beijing, China} 
	\IEEEauthorblockA{\IEEEauthorrefmark{5}School of Integrated Circuits, Southeast University, Nanjing, China}

     \IEEEauthorblockA{\IEEEauthorrefmark{*}Corresponding Author: Li Du \quad Email: ldu@nju.edu.cn}
}

\maketitle

\begin{abstract}
Recent advancements have demonstrated the significant potential of large language models (LLMs) in analog circuit design. Nevertheless, testbench construction for analog circuits remains manual, creating a critical bottleneck in achieving fully automated design processes. Particularly when replicating circuit designs from academic papers, manual Testbench construction demands time-intensive implementation and frequent adjustments, which fails to address the dynamic diversity and flexibility requirements for automation. AnalogTester tackles automated analog design challenges through an LLM-powered pipeline: a) domain-knowledge integration, b) paper information extraction, c) simulation scheme synthesis, and d) testbench code generation with Tsinghua Electronic Design (TED). AnalogTester has demonstrated automated Testbench generation capabilities for three fundamental analog circuit types: operational amplifiers (op-amps), bandgap references (BGRs), and low-dropout regulators (LDOs), while maintaining a scalable framework for adaptation to broader circuit topologies. Furthermore, AnalogTester can generate circuit knowledge data and TED code corpus, establishing fundamental training datasets for LLM specialization in analog circuit design automation.
\end{abstract}


\section{Introduction}
A testbench serves as a critical verification environment in circuit design, comprising three essential components: excitation signal generators, a device under test (DUT) instance, and result monitoring/comparison modules\cite{kitchenStimulusGenerationConstrained2007}. Automated testbench construction enables closed-loop design-simulation-optimization workflows, dramatically accelerating automated circuit design space exploration\cite{qiuAutoBenchAutomaticTestbench2024a}.

The testbenches for digital circuits and analog circuits are different. Testbenches in digital IC design usually pay more attention to the verification of logical functions. Various digital signal patterns, such as random data and specific test vectors, can be used to verify the logical correctness of digital circuits through the use of scripting languages or hardware description languages. Many breakthroughs have been made in the automatic generation of testbenches for digital circuits. Large language models (LLMs) based testbench construction\cite{qiuAutoBenchAutomaticTestbench2024a}, RTL generation\cite{huUVLLMAutomatedUniversal2024,InvitedPaperVerilogEval,changChipGPTHowFar2023,blockloveChipChatChallengesOpportunities2023a,thakurAutoChipAutomatingHDL2024}, formal verification\cite{orenes2023using}, stimuli generation\cite{zhang2023llm4dv} are very impressive.

However, analog testbenches demand sophisticated signal sources (e.g., precision-tuned sine waves) and complex setups adapting to diverse performance metrics (gain, offset, noise, etc.). This domain-specific complexity results in three fundamental limitations: 1) poor reusability across different circuit topologies, 2) absence of universal verification templates, and 3) requirement for manual calibration of simulation parameters. Current industrial practice still relies heavily on experienced designers to create customized testbenches through iterative refinement.

The proposal of the Tsinghua Electronic Design (TED)\cite{yeTEDPythonBasedAnalog2023}, a python-based analog hardware design environment, enables code-driven full-process design implementation, making it an optimal solution for analog circuit testbench construction. Leveraging extensive python corpus training\cite{chenEvaluatingLargeLanguage2021}, LLMs demonstrate superior code comprehension and generation capabilities, showing potential for TED code generation. Furthermore, LLMs demonstrates a broad understanding of analog circuit knowledge, enabling systematic interpretation of analog circuits tasks, and even serving as knowledge bases for data synthesis\cite{chenArtisanAutomatedOperational2024a,liuLADACLargeLanguage,liuAmpAgentLLMbasedMultiAgent2024}.

However, realizing automatic generation of analog testbench based on LLMs remains challenging. First, LLMs cannot directly perform end-to-end processing of such complex system-level tasks. Second, insufficient TED-specific code corpora and systematic testbench design knowledge significantly limit LLM's accuracy in testbench generation. Additionally, inherent limitations like hallucination phenomena necessitate rigorous verification and iterative refinement of LLMs' outputs to ensure reliability\cite{SurveyHallucinationLarge}.

This paper presents AnalogTester, an automated testbench generation framework for analog integrated circuits, implemented through an LLM-based multi-agent system. The framework achieves full-process automation by using research papers as input to extract circuit specifications and synthesize corresponding testbenches, establishing a foundation for large-scale replication of analog circuit research and flexible design optimization. The main contributions include:

\begin{figure*}[!t]
\centering
\includegraphics[width=\linewidth]{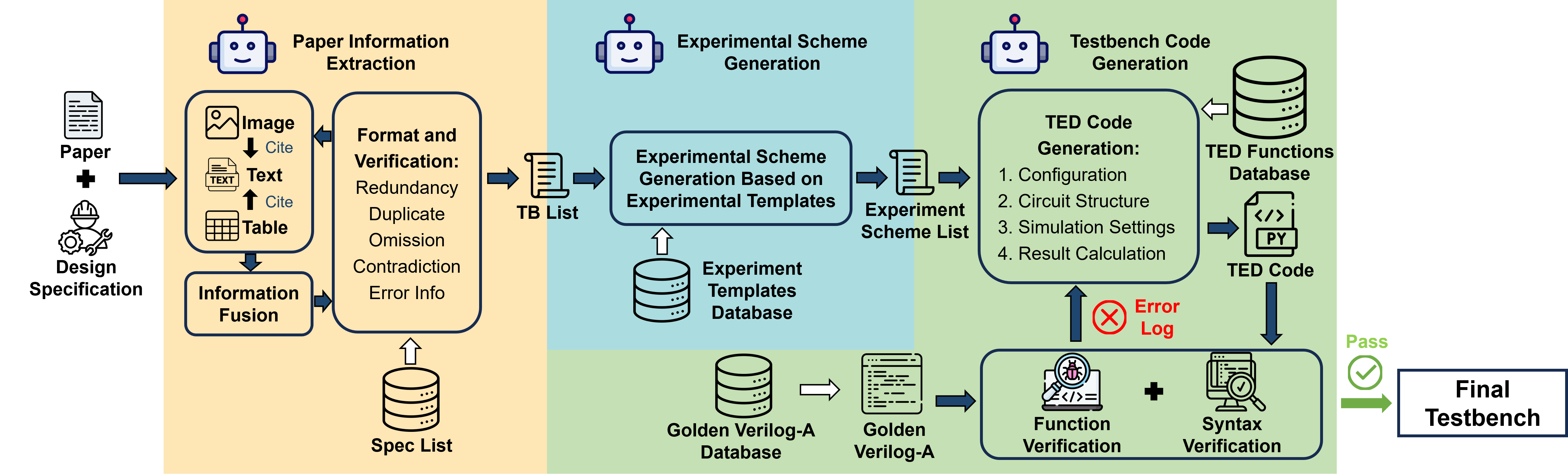}
\caption{The system workflow of AnalogTester.}
\label{fig1}
\end{figure*}

\begin{itemize}
\item A system-level solution for analog circuit testbench generation. The tasks are decomposed into three major modules: paper information extraction, experimental scheme generation, and testbench code generation.
\item We developed a knowledge repository of testbench experimental templates through LLM-driven rapid automation, effectively addressing the lack of systematic comprehension of circuit knowledge in LLMs.
\item The RAG database of TED code is constructed for the first time, significantly enhancing the LLM’s understanding and generation ability for TED.
\item In order to solve the instability problem of LLM output, we propose a large number of iterative error correction frameworks in the process. Combined with Verilog-A, automatic TED syntax and functional verification are realized.
\item We have implemented automatic generation of testbenches for three basic circuit types: operated amplifiers(op-amps), bandgap references (BGRs), and low-dropout regulators (LDOs). Meanwhile, AnalogTester has the potential for rapid expansion.
\end{itemize}

\section{Background}
\subsection{Verilog-A and TED}
Verilog-A is a behavioral-level hardware description language (HDL) for analog/mixed-signal (AMS) design. It enables both ideal transfer characterization and enhanced simulation accuracy through non-ideal parameter integration (e.g., process variations, temperature drift). Practically, it serves as an efficient alternative to transistor-level implementations for system-level verification\cite{588533}.

TED is an embedded domain-specific language and accompanying design environment\cite{yeTEDPythonBasedAnalog2023} built using Python and aimed at enabling codified analog design. TED achieves end-to-end automation spanning from circuit schematic description to physical layout implementation. At the infrastructure level, TED maintains compatibility with conventional EDA tools (e.g., Cadence Virtuoso), while its technology-generic code abstraction layer enables cross-process-node adaptability\cite{wangAutomatedGenerationProcedure2024}. Through the systematic integration of code abstraction mechanisms and automated engines, TED delivers a highly flexible and reusable solution for analog integrated circuit design.

The adoption of TED language for analog circuit testbench construction enables seamless interaction between testbenches and upstream/downstream tasks. Concurrently, replacing transistor-level implementations with golden Verilog-A modules for testbench validation can significantly accelerate automated design and verification workflows.

\subsection{LLM for Analog Circuit}
The application of LLMs in analog circuit design remains an evolving research frontier. Current implementations have demonstrated foundational capabilities in tasks, including topology generation\cite{chenArtisanAutomatedOperational2024a,changLaMAGICLanguageModelbasedTopology2024}, sizing optimization\cite{yinADOLLMAnalogDesign2024}, netlist synthesis\cite{bhandari2025masalachailargescalespicenetlist}, and design assistance\cite{liuLADACLargeLanguage}, showcasing LLMs' potential to comprehend and contribute to analog circuit development. However, existing "automated design" frameworks still rely on manually preconfigured testbenches, which constrains their applicability to specific circuit types like operational amplifiers and prevents broader generalization across analog domains.

A critical bottleneck arises from the stark data scarcity in analog circuit design compared to the massive parameter scale of LLMs (typically exceeding billions of parameters). This data paucity fundamentally limits both the training of domain-specific LLMs and their generalizability. Consequently, adopting a multi-agent framework to construct automated workflows demonstrates significant practical value: it not only dramatically accelerates analog circuit design processes but also systematically generates structured circuit knowledge datasets, thereby establishing a foundational corpus for training domain-adapted AI models.

\section{AnalogTester: Automatic Testbench Generation for Analog Circuits}
 To address the challenges proposed in Section \Rmnum{1}, this section elaborates on the technical details of AnalogTester. As shown in Fig. \ref{fig1}, AnalogTester primarily consists of three core components: Information Extraction Agent, Experimental Scheme Generation Agent, and Testbench Code Generation Agent. Unless otherwise specified, all core functions shown in Fig. \ref{fig1} are implemented by LLM multi-agent system based on OpenAI GPT-4o\cite{OpenAIPlatform}.

The Information Extraction Agent takes research papers and user requirements as input, extracting simulation information to generate a list of target testbenches. Subsequently, the Experimental Scheme Generation Agent leverages a pre-constructed circuit knowledge to emulate human expert behavior in designing specific testbench solutions based on the generated list. Finally, the Testbench Code Generation Agent functions as a TED code expert system, responsible for producing TED code of the testbenches.

\subsection{Information Extraction}
In the yellow area of Fig. \ref{fig1}, AnalogTester implements the extraction and verification of multimodal simulation information in the paper, automatically establishing a task list for testbench construction.

\textbf{Multimodal Information Extraction: }Extract the simulation information of text, pictures, and tables from papers in markdown format as input. The extraction of three types of modal information is each handled by an LLM agent.
\begin{itemize}
\item Text extraction: The simulation information described in words is extracted from the target paper, covering simulation parameter settings, experimental conditions, expectations indicators and other contents. The extraction results also include the name of the circuit under test, as well as the table and image cited by each experiment.
\item Picture extraction: Screen different types of experimental result graphs through visual multimodal LLM and design prompts for them separately. Then, conduct in-depth processing on the horizontal and vertical coordinates, legends, text annotations, curve trends, etc. of the pictures.
\item Table extraction: Mainly for the comparison work tables and experimental result tables, summarize and integrate the information in the tables into the total results.
\end{itemize}

\textbf{Information Fusion: }After the extraction, the information of each modality has been transformed into natural language. In a well-written academic paper, the contents in the tables and pictures should be accurately cited and explained in the text information. Therefore, the text information is the main source for the testbench, while the picture and table information serve as supplements. Add picture and table information to text information based on citation relationships.

\textbf{Information Format and Optimization: }Based on the self-reflection mechanism of the LLMs, an agent expert for checking and verifying the extracted information is designed. \textbf{Spec list} is used to indicate the basic simulation items that must be included in the current circuit type. Through the completeness evaluation, omissions, redundancies, or inconsistencies in the data are identified, with error feedback enabling iterative refinement to enhance information extraction quality. This integrity verification ensures formatted outputs, demonstrating significant improvement in completeness and accuracy compared to the direct output without verification.

\begin{figure}[htbp]
\centerline{\includegraphics[width=1.0\columnwidth]{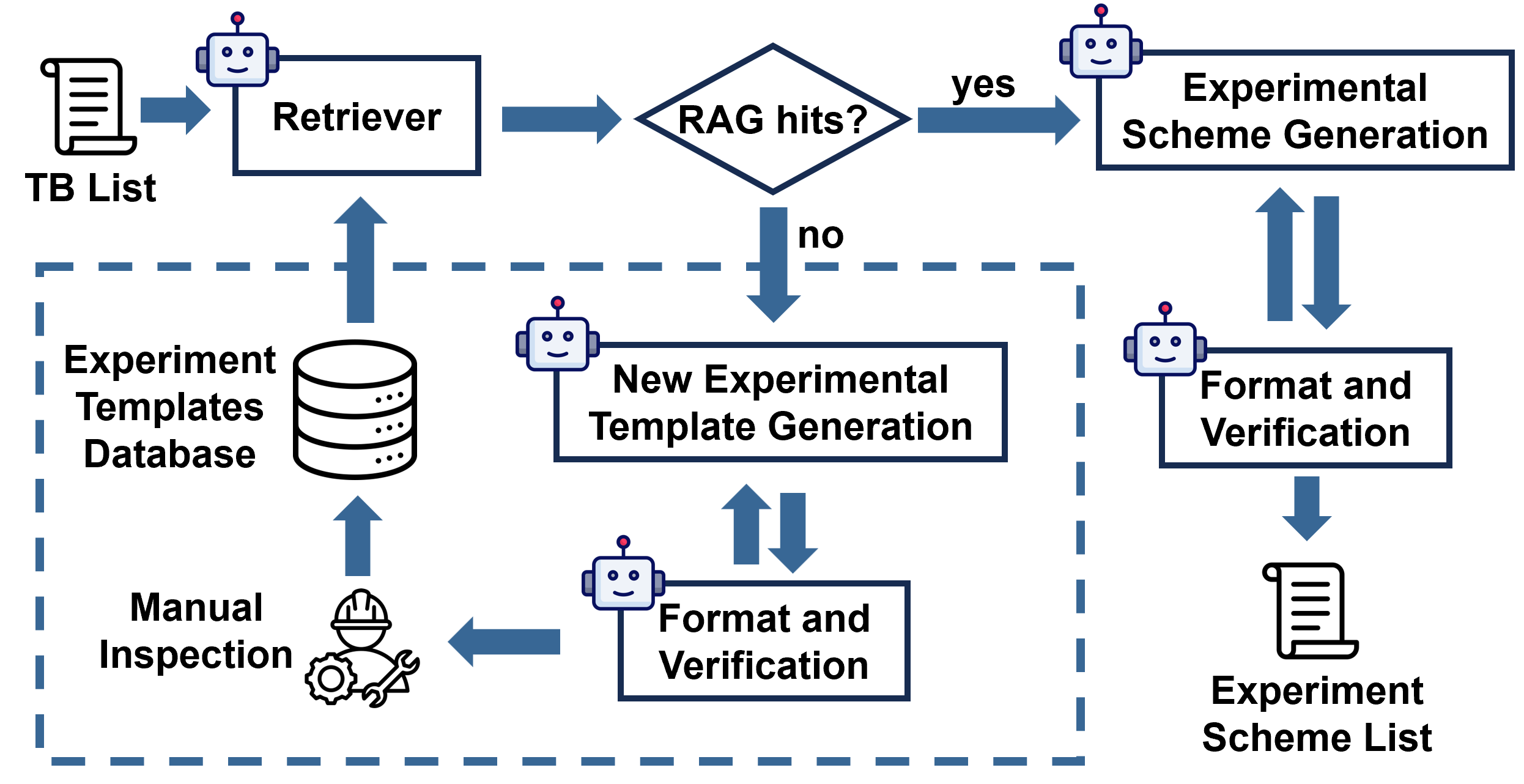}}
\caption{Experimental Scheme and Experimental Template Generation Framework.}
\label{fig2}
\end{figure}

\begin{table}[!ht]
\centering
    \belowrulesep=0pt
\aboverulesep=0pt
    \caption{Basic Performance Metrics Overview in Experimental Templates Database}
    \begin{tabular}{c|p{5cm}}
    \toprule
        	\textbf{Circuit Type} & \textbf{Performance Metrics} \\ 
         \hline
        \textbf{OP} & \multicolumn{0}{m{6cm}}{CMRR, Gain, GBW, GM, ICMR, Noise, Offset, Output swing, PM, Power, PSRR, Slew-Rate, THD}\\  \hline
        \textbf{BGP} & \multicolumn{0}{m{6cm}}{LNR, Load-test, Noise, PSRR, Quiescent Current, Power, Self-startup, TC, Transient Startup} \\  \hline
        \textbf{LDO} & \multicolumn{0}{m{6cm}}{ Current-efficiency, Dropout, Efficiency, LDR, LNR, Maximum Output Current, Output Voltage, Output Voltage Noise, Overshoot, Undershoot, Power, PSRR, Quiescent Current, Settling-time} \\ \bottomrule
    \end{tabular}
    \label{perf_table}
\end{table}

\subsection{Experimental Scheme Generation}
The simulation requirements of analog circuits vary significantly across circuit types and performance metrics, posing challenges in developing standardized analog testbenches. While LLMs demonstrate broad knowledge coverage for basic analog circuit testbenches, their capabilities remain unsystematic. Additionally, the absence of domain-specific prior knowledge for TED code generation hinders direct translation of experimental requirements into simulation code.

AnalogTester introduces a task decomposition methodology, decoupling circuit-related tasks from code-generation tasks and separating knowledge construction from comprehension generation. This approach significantly reduces the complexity of handling intricate problem-solving processes.

As illustrated in the blue area of Fig. \ref{fig1}, we developed an experimental scheme template repository to facilitate automated experimental design generation. Each template comprises four standardized components: foundational PVT (Process, Voltage, Temperature) parameters, testbench circuit architecture, simulation types, and result calculation methodologies. The current implementation encompasses 36 experimental templates targeting the most prevalent performance metrics for three fundamental analog circuits: op-amp, BGR, and LDO, which are shown in Table \ref{perf_table}.

The generation of experimental schemes based on templates and the method of experimental templates database augmentation based on the construction of new templates are discussed in detail in Fig. \ref{fig2}. AnalogTester sequentially processes testbench task list through template repository retrieval. Upon successful template matching, it employs Retrieval-Augmented Generation (RAG) to refine the template by integrating specific experimental conditions, configuration parameters, and target metrics from the task specification, thereby generating optimized experimental schemes.

For unmatched queries, AnalogTester initiates a new template generation agent. This process utilizes prompts to distil domain knowledge from GPT4-o1 in constrained format, guided by circuit name and target performance metric. The system subsequently enforces rigorous format compliance verification through an auxiliary LLM, implementing critical evaluation and validation of initial outputs via iterative feedback refinement. Finalized templates undergo human expert validation before repository integration.

\begin{figure}[ht]
\centerline{\includegraphics[width=1.0\columnwidth]{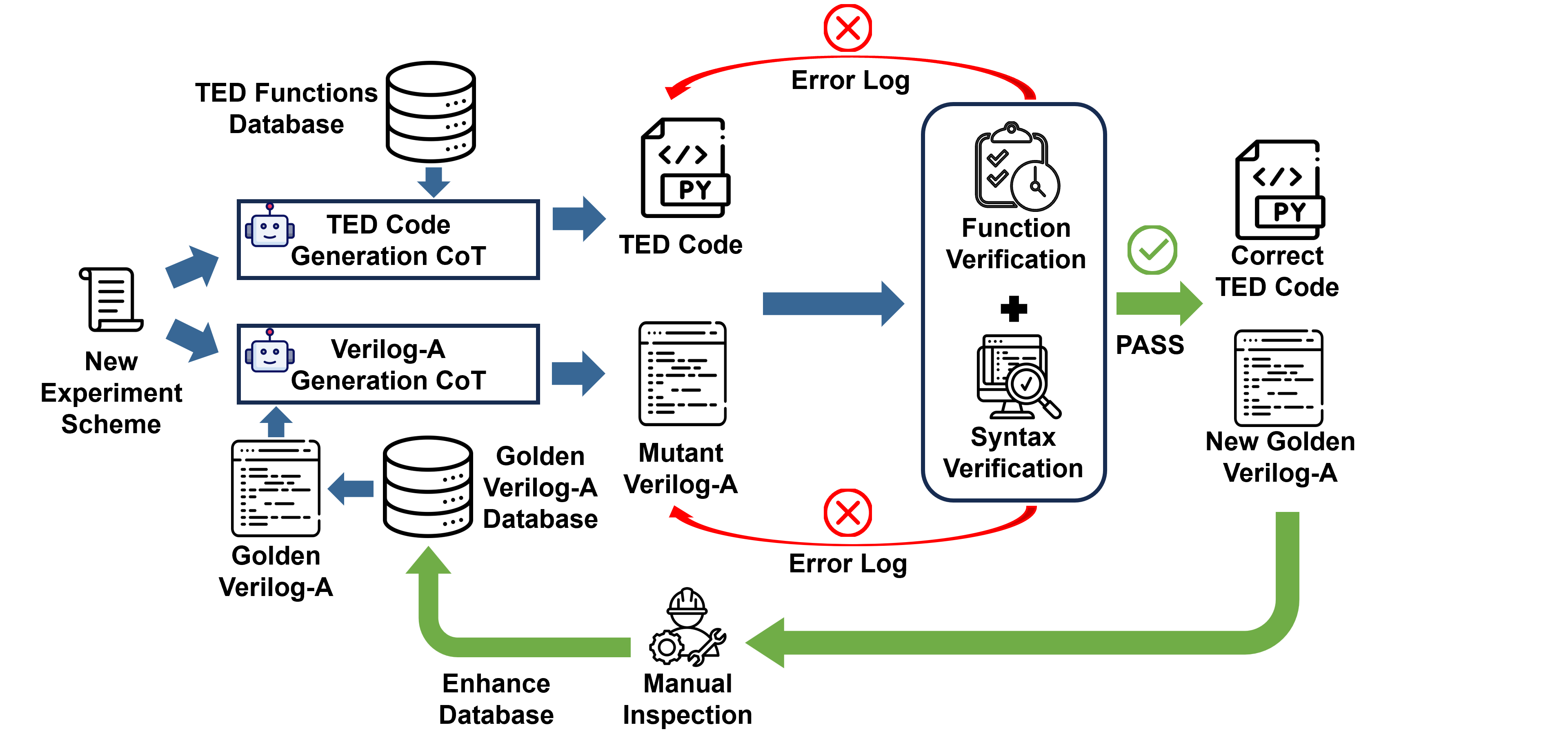}}
\caption{Collaborative Generation and Verification of Testbench and Verilog-A.}
\label{fig3}
\end{figure}

\subsection{Testbench Code Generation}
We have manually curated a TED function database. This database provides comprehensive coverage of 72 foundational functions essential for testbench construction and front-end/back-end design implementation, methodically categorized as: 16 node definition functions, 9 simulation functions, 30 data processing and computation modules, and 17 layout functions. This not only provides knowledge for the construction of testbench but also accumulates corpora for the future training of LLMs for TED. It is the first basic function dataset of TED. The TED function database explains the names, functions, and usage of TED functions, which can enable LLM to acquire TED generation capabilities. An example of TED function database is shown in Fig. \ref{fig6}.

At the same time, we have also modelled the Verilog-A models of op-amp, BGR and LDO for the performance metrics included in the experimental scheme template database mentioned earlier. These golden Verilog-A are used to verify the generated TED code. 

As shown in the green area of Fig. \ref{fig1}, the TED code iterative optimization mechanism is as follows: 

\begin{itemize}
    \item According to the experimental scheme, decompose the testbench code generation task into CoT, and construct TED for each performance metric separately. The code is generated based on RAG from the TED function database.
    \item Replace the DUT with the Golden Verilog-A, execute the TED code, and conduct syntax and functionality verification.
    \item If the checks are passed and the results match the target performance of the golden Verilog-A, generate the final testbench; otherwise, capture the execution errors and feed them back to the TED-generated LLM, enabling it to automatically generate the repair code. 
\end{itemize}

When the experimental scheme is built on a newly generated template, the golden Verilog-A in the database must be re-modelled for newly added performance metrics to enable system extensibility. Fig. \ref{fig3} demonstrates the extended generation and mutual verification mechanism between TED and Verilog-A. The LLM CoT generates mutant Verilog-A by integrating new performance metrics with the existing golden Verilog-A, which are then jointly verified with the TED code. Verification is passed only upon correct co-generation of both TED and Verilog-A, thereby enabling the production of validated TED code and expanding the Verilog-A database. To ensure the correctness of the Golden VA database, manual verification will be introduced before the end of the process.

\begin{table}[!t]
\centering
\belowrulesep=0pt
\aboverulesep=0pt
    \caption{Simulation Results of the OTA}
    \label{simu_result}
    \begin{tabular}{c|ccc}
        \toprule
        \textbf{\makecell{Metrics}} & \textbf{\makecell{TED}} & \textbf{\makecell{Cadence Virtuoso}}\\ \hline
        Open loop gain & 71.1 dB & 71.1 dB\\
        UGW & 33.1 kHz & 33.1 kHz\\
        Phase margin & 75.3° & 75.3°\\
        Maximum swing & 0-0.60 V & 0-0.60 V\\
        Linear swing & 0.03-0.58 V & 0.03-0.58 V\\
        Slew-rate & 30.9 V/ms & 30.9 V/ms\\
        CMRR@100Hz & 85.7 dB & 85.7 dB\\
        PSRR@100Hz & 63.0 dB & 63.0 dB\\
        Power consumption & 365 nW & 365 nW\\
        THD@520mV\(_{pp}\) & 0.31\% & 0.31\%\\
        Minimum supply & 600 mV & 600 mV\\
        \bottomrule
\end{tabular}
\end{table}

\section{Implement of AnalogTester}
We take the op-amp published by Luís H. C. Ferreira in 2007\cite{ferreiraUltraLowVoltageUltraLowPowerCMOS2007} as an example to demonstrate the specific generation effects of AnalogTester. We replicated the circuit using TSMC 65nm technology and redesigned the transistor size. In the article, the op-amp was simulated on 11 performance indicators, which is shown in Table \ref{simu_result}. Table \ref{simu_result} shows that the ted testbenches generated by AnalogTester have good consistency with the manually constructed testbenches in Cadence Virtuoso.

\begin{figure}[ht]
\centerline{\includegraphics[width=1.0\columnwidth]{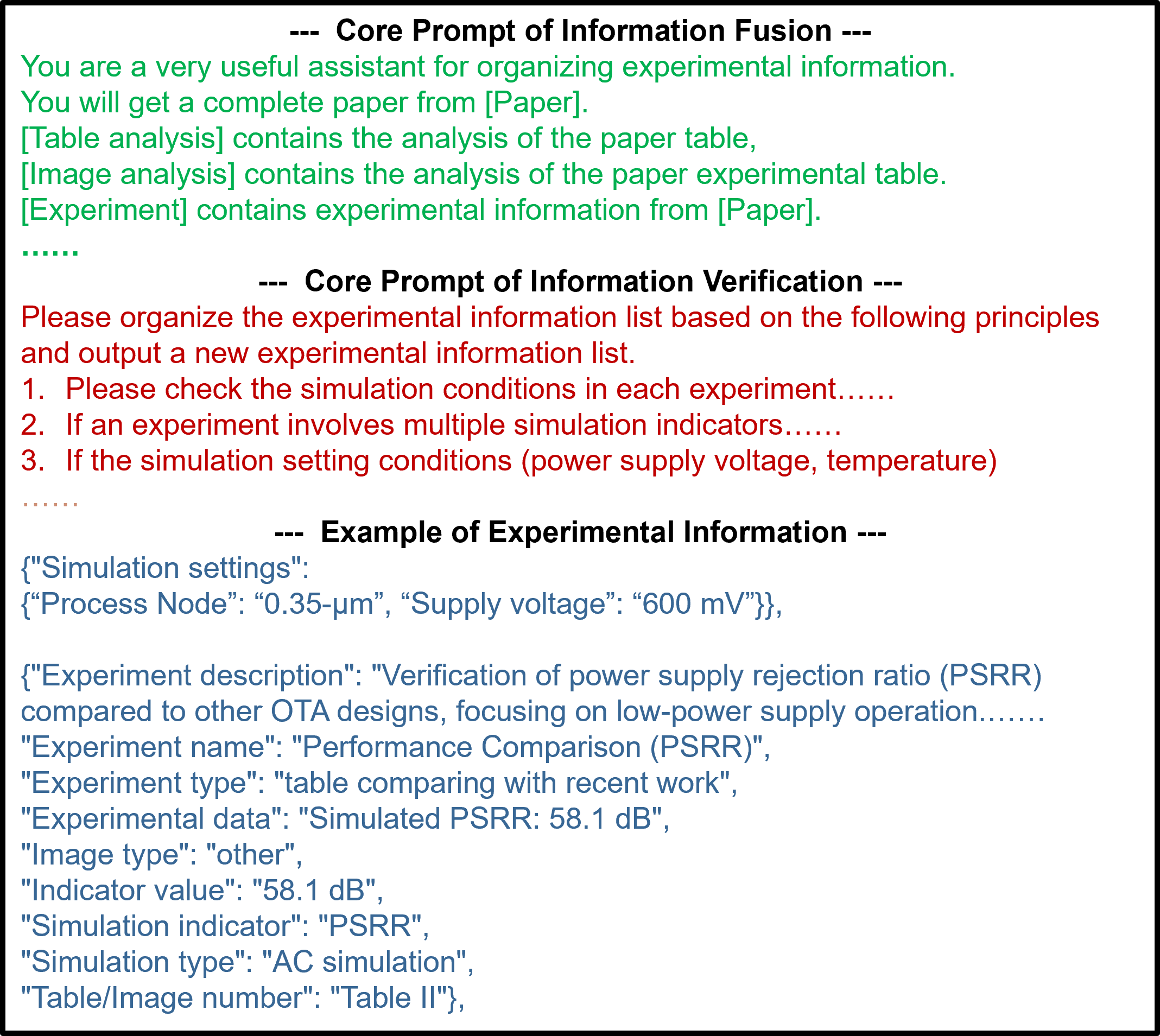}}
\caption{Core prompt and example of results of information extraction agent.}
\label{fig4}
\end{figure}

\begin{figure}[ht]
\centerline{\includegraphics[width=1.0\columnwidth]{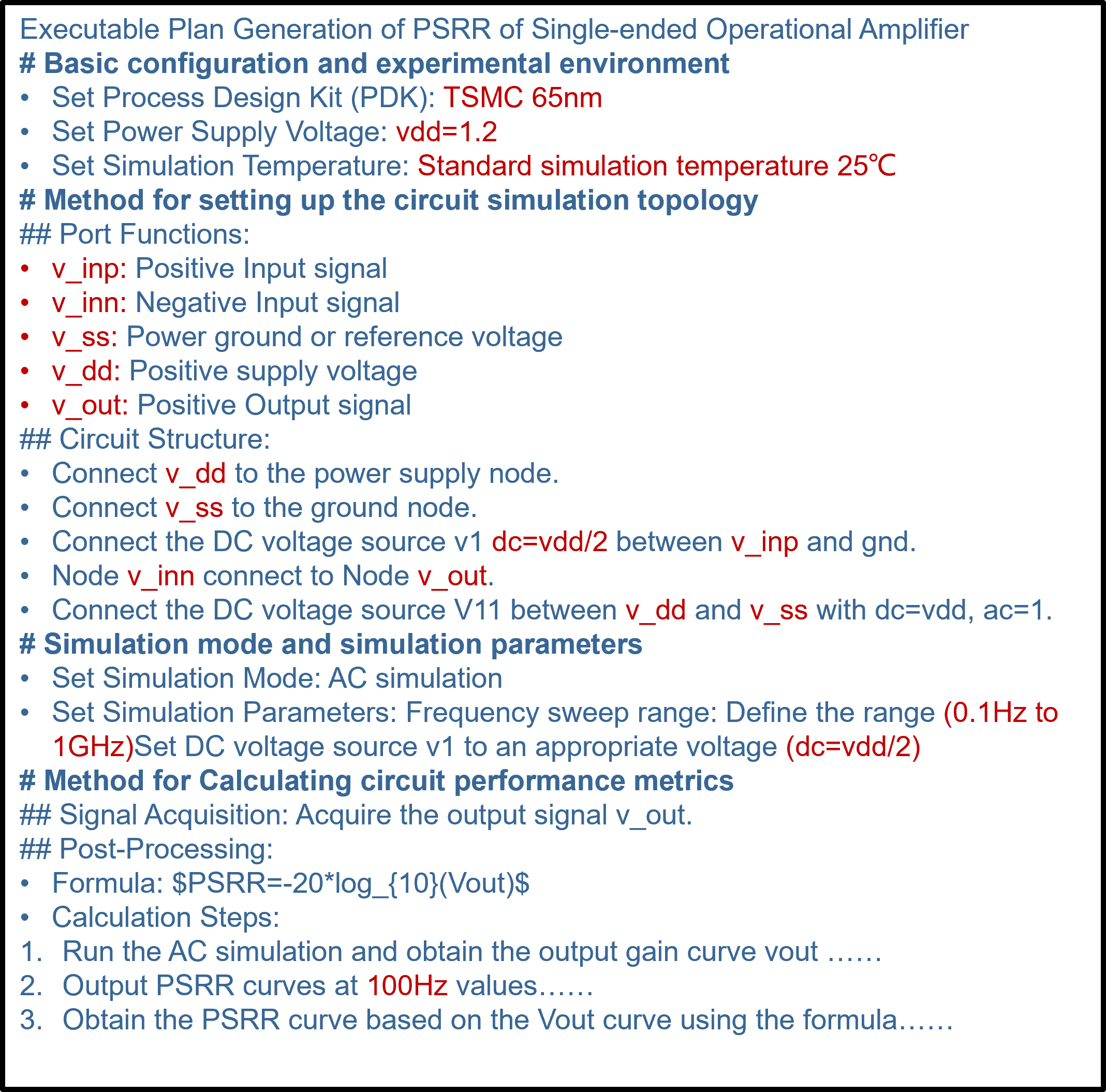}}
\caption{An example of experimental scheme generation result.}
\label{fig5}
\end{figure}

AnalogTester organizes experimental information into clear structured text for subsequent processes. The core prompts of generation and verification and an extraction results example of the paper information extraction agent are shown in Fig. \ref{fig4}. The extracted experimental information is in JSON format, with each experiment independently maintaining its own fields. After manual confirmation, AnalogTester performs RAG from the experimental templates database, then integrates extracted experimental conditions and settings to generate detailed testbench schemes step-by-step. Fig. \ref{fig5} shows the testbench of power supply rejection ratio (PSRR) as an example, with the blue part representing the main content of the experimental template and the red part representing the experimental scheme generated based on the template.

\begin{figure}[ht]
\centerline{\includegraphics[width=1.0\columnwidth]{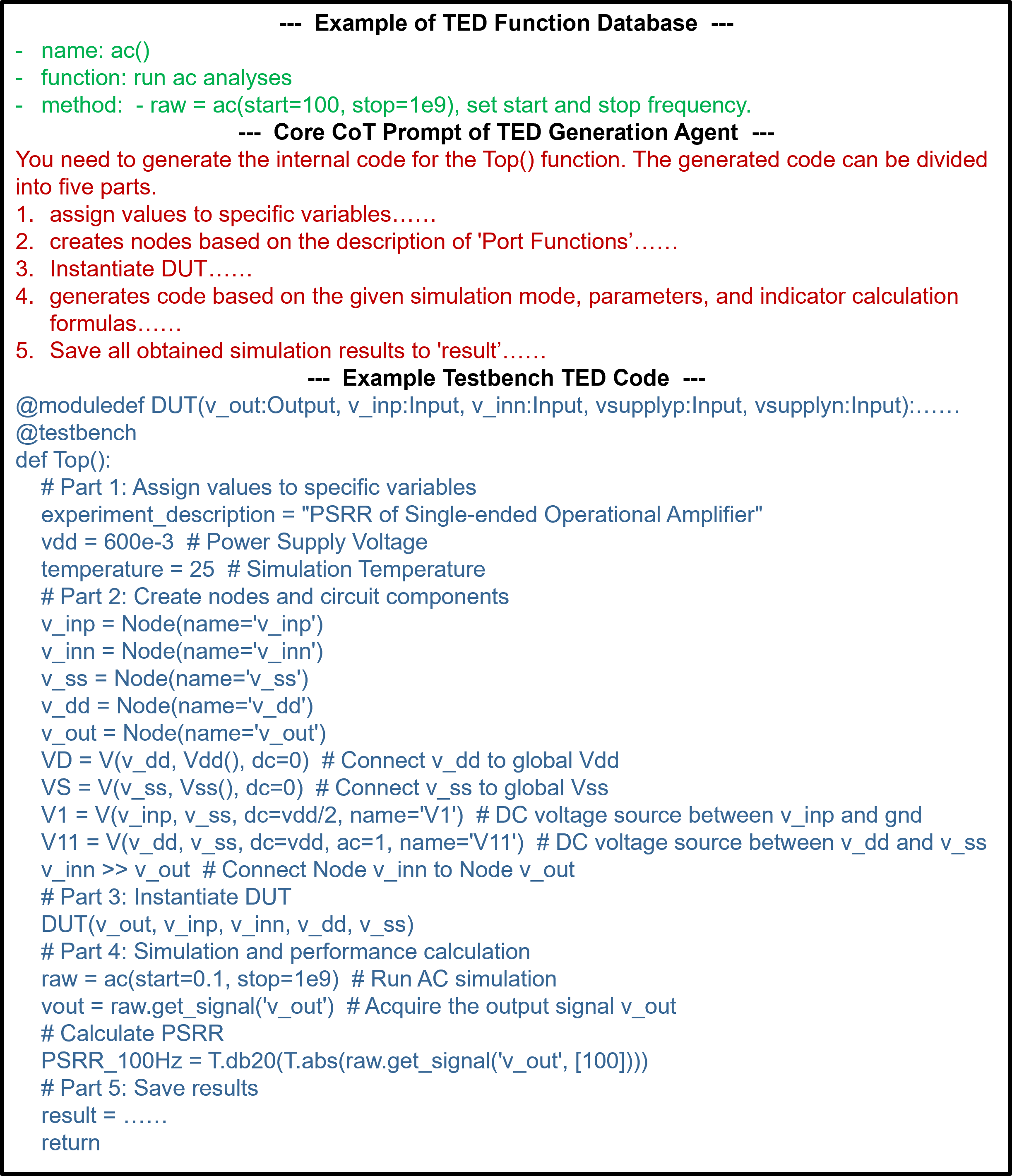}}
\caption{Core prompt and example of results of TED code generation agent.}
\label{fig6}
\end{figure}

Finally, AnalogTester generates TED code based on the experimental schemes. The example of TED function data, the main generation prompt and the sample TED code of PSRR are shown in Fig. \ref{fig6}. For TED code generation agent, the circuit knowledge comes from experimental schemes, and TED code usage knowledge comes from TED function databases. It only needs to generate TED code correctly based on LLMs' understanding ability. In the example code, DUT is used to encapsulate Verilog-A or circuit netlist, while testbench code is encapsulated in the \textbf{top()} function.

\section{Evaluation and Experiments of AnalogTester}
\subsection{Basic Evaluation Method}
We constructed a benchmark comprising 8 research papers (3 op-amps, 3 BGRs, and 2 LDOs)\cite{ferreiraUltraLowVoltageUltraLowPowerCMOS2007,wang05VOperationalTransconductance2006,lauMillercompensatedAmplifierGmboosting2015,colomboCMOS253Ppmdeg2012,giustolisiLowvoltageLowpowerVoltage2003,ueno300NW152009,jiang65nmCMOSLow2018,el-nozahiHighPSRLow2010} to systematically evaluate AnalogTester's generation accuracy. Given the functional independence of AnalogTester's three core agent frameworks, we implemented separate evaluation protocols. To reproduce these papers, 63 simulations are required, which are simulated and discussed in these papers. This means that AnalogTester needs to extract information for these 63 experiments and generate experimental schemes and codes for them separately. Among the 63 experiments, 23 involved op-amps, 18 focused on BGRs, and 22 were dedicated to LDOs.

The success rates and time consumption are shown in Table \ref{tab3}. We manually ensure that the inputs upstream task for each task is correct, decoupling the success rates between tasks. The \textbf{Total} indicates the overall success rate for each task. All three tasks achieved accuracy rates above 80\%, demonstrating robust usability. \textbf{Avg. Time} represents the average time required to process a single experiment within a task. Despite frequent slowdowns caused by rate limiting in LLMs, the rapid generation speed of AnalogTester enables the full process of constructing a paper’s testbench to be completed in under an hour. Furthermore, the processing speed exhibits potential for further optimization via parallel processing mechanisms in the future.

All agents are driven by the GPT-4o model, and the temperature is set to 0. The processor of the experimental platform is Intel® Xeon® CPU E5-2698 v4 @ 2.20GHz. The large language model service is invoked through the API-key and does not consume local computing power.

\begin{table}[!t]
\setlength{\extrarowheight}{1ex}
\centering
\belowrulesep=0pt
\aboverulesep=0pt
    \caption{Accuracy of Benchmark Testing Tasks}
    \label{tab3}
    \begin{tabular}{c|ccccc}
        \toprule
        \textbf{\makecell{Task}} & \textbf{\makecell{op-amp}} & \textbf{\makecell{BGR}} & \textbf{\makecell{LDO}} & \textbf{\makecell{Total}} & \textbf{\makecell{Avg. Time}}\\ \hline
        Task1 & 21/23  & 15/18 & 17/22 & 84\% & 123 sec\\
        Task2 & 20/23 & 17/18 & 17/22 & 86\% & 165 sec\\
        Task3 & 21/23 & 17/18 & 18/22 & 89\% & 126 sec\\
        \bottomrule
        \multicolumn{6}{l}{Task1: Information Extraction.}\\
        \multicolumn{6}{l}{Task2: Experiment Scheme Generation.}\\
        \multicolumn{6}{l}{Task3: Testbench Code Generation.}\\
\end{tabular}
\end{table}

\subsection{Ablation Experiment}
\subsubsection{Impact of the Priori Knowledge}
\begin{figure}[ht]
\centering
\centerline{\includegraphics[width=1\columnwidth]{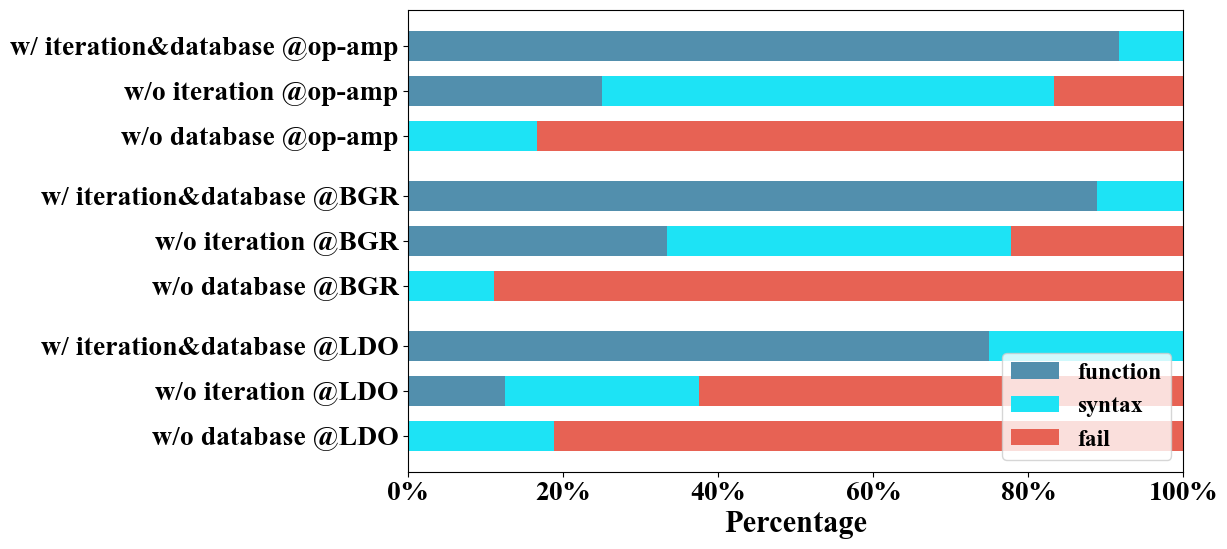}}
\caption{Ablation experiment on TED generation task with maximum iteration times of 9.}
\label{ablation_iter&database}
\end{figure}

In Fig. \ref{ablation_iter&database}, we demonstrate the influence of the TED function knowledge base on the task of generating Testbench code. In this part of the experiment, the retrieval enhancement from the knowledge base was removed within the normal framework of AnalogTester, and the GPT-4o model was still used as the underlying model.

We conducted TED code generation using 37 predefined templates(12 of op-amps, 9 of BGRs, 16 of LDOs). The control group employed AnalogTester (with iterative refinement and TED Function database), configured for a maximum of 9 iterations. Group \textbf{w/o iteration} used single-pass generation, while Group \textbf{w/o database} removed the TED Function database and relied solely on RAG with TED documentation. Results showed AnalogTester achieved a \textbf{100\%} syntax pass rate and over 80\% functional check success rate. Single-pass generation yielded below 22\% functional success, and exclusion of the TED Function database caused complete failure in functional verification.

\subsubsection{Impact of Model Capability and impact of the AnalogTester Framework}
In order to analyze the impact of the model capabilities on the generation success rate of AnalogTester, we carried out a comparative experiment on replacing the underlying model of AnalogTester. We replaced the GPT-4o model used at the underlying layer of AnalogTester with GPT-o3mini and DeepSeek-R1, and compared the accuracy rates of various tasks on an OTA papers\cite{ferreiraUltraLowVoltageUltraLowPowerCMOS2007} in the basic evaluation experiment. It should be noted that we still used GPT-4o for the image information processing in Deepseek-R1 and GPT-o3mini. We also designed a comparative experiment \textbf{CPT-4o without AnalogTester} using the GPT-4o model: Task 1 applied single-pass information extraction without verification, Task 2 omitted experimental templates and used simple prompt-driven generation, and Task 3 disabled the TED function database. This setup highlights the critical contributions of AnalogTester’s core methodologies across all tasks.

\begin{figure}[ht]
\centering
\centerline{\includegraphics[width=0.85\columnwidth]{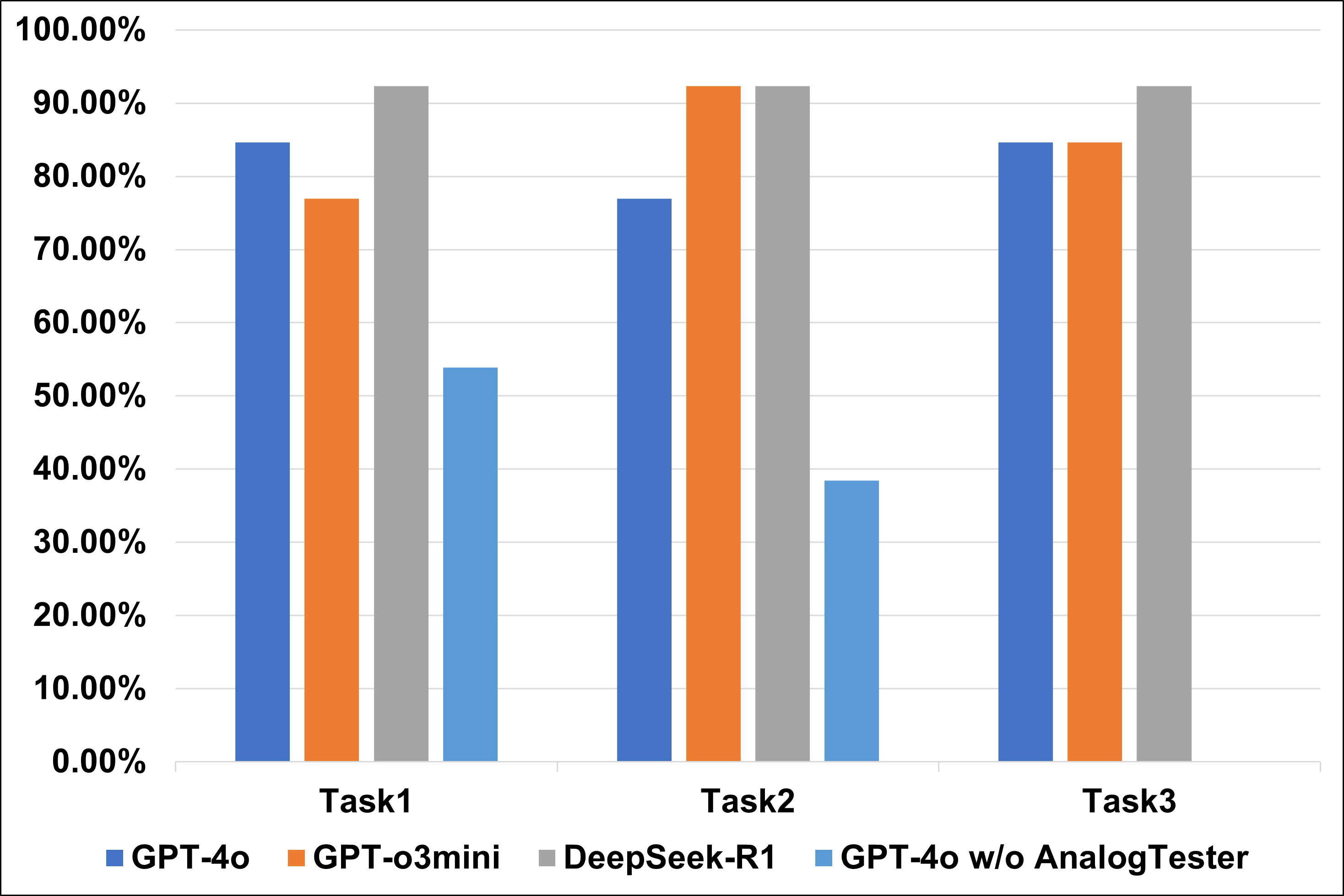}}
\caption{Results of AnalogTester Iteration Framework Ablation Experiment.}
\label{fig8}
\end{figure}

The results are shown in Fig. \ref{fig8}. Given the high overall success rates across tasks, the three models showed no significant performance differences. However, Group \textbf{GPT-4o without AnalogTester} exhibited notably lower success rates, effectively demonstrating the efficacy of AnalogTester's core methodologies in the workflow. Notably, the GPT-o3mini and DeepSeek-R1 required 3–10x longer processing times per task compared to the GPT-4o model.

\section*{Conclusion}
This paper presents AnalogTester, an automatic framework for analog testbench generation based on LLMs. For the first time, the challenges of automatic analog circuit testbench generation have been addressed through methods such as task decomposition, iterative verification, knowledge repository construction, and automated scalability. AnalogTester successfully generates testbenches for op-amps, BGRs, and LDOs, demonstrating scalability for broader analog circuits. This work establishes a foundation for systematic, large-scale analog design automation and optimization.

\bibliographystyle{ieeetr}
\bibliography{main.bbl}

\begin{thebibliography}{10}

\bibitem{kitchenStimulusGenerationConstrained2007}
N.~Kitchen and A.~Kuehlmann, ``Stimulus generation for constrained random
  simulation,'' in {\em 2007 {{IEEE}}/{{ACM International Conference}} on
  {{Computer-Aided Design}}}, pp.~258--265, Nov. 2007.

\bibitem{qiuAutoBenchAutomaticTestbench2024a}
R.~Qiu, G.~L. Zhang, R.~Drechsler, U.~Schlichtmann, and B.~Li, ``{{AutoBench}}:
  {{Automatic Testbench Generation}} and {{Evaluation Using LLMs}} for {{HDL
  Design}},'' in {\em Proceedings of the 2024 {{ACM}}/{{IEEE International
  Symposium}} on {{Machine Learning}} for {{CAD}}}, (Salt Lake City UT USA),
  pp.~1--10, ACM, Sept. 2024.

\bibitem{huUVLLMAutomatedUniversal2024}
Y.~Hu, J.~Ye, K.~Xu, J.~Sun, S.~Zhang, X.~Jiao, D.~Pan, J.~Zhou, N.~Wang,
  W.~Shan, X.~Fang, X.~Wang, N.~Guan, and Z.~Jiang, ``{{UVLLM}}: {{An Automated
  Universal RTL Verification Framework}} using {{LLMs}},'' Nov. 2024.

\bibitem{InvitedPaperVerilogEval}
``Invited {{Paper}}: {{VerilogEval}}: {{Evaluating Large Language Models}} for
  {{Verilog Code Generation}} {\textbar} {{IEEE Conference Publication}}
  {\textbar} {{IEEE Xplore}}.''
  https://ieeexplore.ieee.org/abstract/document/10323812.

\bibitem{changChipGPTHowFar2023}
K.~Chang, Y.~Wang, H.~Ren, M.~Wang, S.~Liang, Y.~Han, H.~Li, and X.~Li,
  ``{{ChipGPT}}: {{How}} far are we from natural language hardware design,''
  June 2023.

\bibitem{blockloveChipChatChallengesOpportunities2023a}
J.~Blocklove, S.~Garg, R.~Karri, and H.~Pearce, ``Chip-{{Chat}}: {{Challenges}}
  and {{Opportunities}} in {{Conversational Hardware Design}},'' in {\em 2023
  {{ACM}}/{{IEEE}} 5th {{Workshop}} on {{Machine Learning}} for {{CAD}}
  ({{MLCAD}})}, pp.~1--6, Sept. 2023.

\bibitem{thakurAutoChipAutomatingHDL2024}
S.~Thakur, J.~Blocklove, H.~Pearce, B.~Tan, S.~Garg, and R.~Karri,
  ``{{AutoChip}}: {{Automating HDL Generation Using LLM Feedback}},'' June
  2024.

\bibitem{orenes2023using}
M.~Orenes-Vera, M.~Martonosi, and D.~Wentzlaff, ``Using llms to facilitate
  formal verification of rtl,'' {\em arXiv preprint arXiv:2309.09437}, 2023.

\bibitem{zhang2023llm4dv}
Z.~Zhang, G.~Chadwick, H.~McNally, Y.~Zhao, and R.~Mullins, ``Llm4dv: Using
  large language models for hardware test stimuli generation,'' {\em arXiv
  preprint arXiv:2310.04535}, 2023.

\bibitem{yeTEDPythonBasedAnalog2023}
Z.~Ye, Z.~Wang, J.~Xin, Y.~Wang, Q.~Qin, C.~Chai, Y.~Lu, J.~Hao, J.~Xiao, and
  Y.~Wang, ``{{TED}}: {{A Python-Based Analog Design Environment}} for {{Agile
  Circuit Development}},'' in {\em 2023 {{International Symposium}} of
  {{Electronics Design Automation}} ({{ISEDA}})}, pp.~5--10, May 2023.

\bibitem{chenEvaluatingLargeLanguage2021}
M.~Chen, J.~Tworek, H.~Jun, Q.~Yuan, Pinto, {\em et~al.}, ``Evaluating {{Large
  Language Models Trained}} on {{Code}},'' July 2021.

\bibitem{chenArtisanAutomatedOperational2024a}
Z.~Chen, J.~Huang, Y.~Liu, F.~Yang, L.~Shang, D.~Zhou, and X.~Zeng, ``Artisan:
  {{Automated Operational Amplifier Design}} via {{Domain-specific Large
  Language Model}},'' in {\em Proceedings of the 61st {{ACM}}/{{IEEE Design
  Automation Conference}}}, (San Francisco CA USA), pp.~1--6, ACM, June 2024.

\bibitem{liuLADACLargeLanguage}
C.~Liu, Y.~Liu, Y.~Du, and L.~Du, ``{{LADAC}}: {{Large Language Model-driven
  Auto-Designer}} for {{Analog Circuits}},''

\bibitem{liuAmpAgentLLMbasedMultiAgent2024}
C.~Liu, W.~Chen, A.~Peng, Y.~Du, L.~Du, and J.~Yang, ``{{AmpAgent}}: {{An
  LLM-based Multi-Agent System}} for {{Multi-stage Amplifier Schematic Design}}
  from {{Literature}} for {{Process}} and {{Performance Porting}},'' Sept.
  2024.

\bibitem{SurveyHallucinationLarge}
``A {{Survey}} on {{Hallucination}} in {{Large Language Models}}:
  {{Principles}}, {{Taxonomy}}, {{Challenges}}, and {{Open Questions}}
  {\textbar} {{ACM Transactions}} on {{Information Systems}}.''
  https://dl.acm.org/doi/abs/10.1145/3703155.

\bibitem{588533}
I.~Miller, D.~FitzPatrick, and R.~Aisola, ``Analog design with verilog-a,'' in
  {\em Proceedings of Meeting on Verilog HDL (IVC/VIUF'97)}, pp.~64--68, 1997.

\bibitem{wangAutomatedGenerationProcedure2024}
Y.~Wang, Q.~Wu, Y.~Wang, Q.~Qin, J.~Hao, C.~Chai, Y.~Lu, J.~Huang, L.~Li, and
  Z.~Ye, ``Automated {{Generation Procedure}} for {{Fully Differential Op-Amp
  Using TED}},'' in {\em 2024 2nd {{International Symposium}} of {{Electronics
  Design Automation}} ({{ISEDA}})}, pp.~100--105, May 2024.

\bibitem{changLaMAGICLanguageModelbasedTopology2024}
C.-C. Chang, Y.~Shen, S.~Fan, J.~Li, S.~Zhang, N.~Cao, Y.~Chen, and X.~Zhang,
  ``{{LaMAGIC}}: {{Language-Model-based Topology Generation}} for {{Analog
  Integrated Circuits}},'' Aug. 2024.

\bibitem{yinADOLLMAnalogDesign2024}
Y.~Yin, Y.~Wang, B.~Xu, and P.~Li, ``{{ADO-LLM}}: {{Analog Design Bayesian
  Optimization}} with {{In-Context Learning}} of {{Large Language Models}},''
  June 2024.

\bibitem{bhandari2025masalachailargescalespicenetlist}
J.~Bhandari, V.~Bhat, Y.~He, S.~Garg, H.~Rahmani, and R.~Karri, ``Masala-chai:
  A large-scale spice netlist dataset for analog circuits by harnessing ai,''
  2025.

\bibitem{OpenAIPlatform}
``{{OpenAI Platform}}.'' https://platform.openai.com.

\bibitem{ferreiraUltraLowVoltageUltraLowPowerCMOS2007}
L.~H.~C. Ferreira, T.~C. Pimenta, and R.~L. Moreno, ``An {{Ultra-Low-Voltage
  Ultra-Low-Power CMOS Miller OTA With Rail-to-Rail Input}}/{{Output Swing}},''
  {\em IEEE Transactions on Circuits and Systems II: Express Briefs}, vol.~54,
  pp.~843--847, Oct. 2007.

\bibitem{wang05VOperationalTransconductance2006}
H.~Wang and Q.~Ye, ``0.5-{{V}} operational transconductance amplifier for
  {{CMOS}} bandgap reference application,'' in {\em 2006 8th {{International
  Conference}} on {{Solid-State}} and {{Integrated Circuit Technology
  Proceedings}}}, pp.~1705--1707, Oct. 2006.

\bibitem{lauMillercompensatedAmplifierGmboosting2015}
M.~W. Lau, M.~Ho, K.~H. Mak, S.~Bu, K.~N. Leung, and W.~L. Goh, ``A
  {{Miller-compensated}} amplifier with {{Gm-boosting}},'' in {\em {{TENCON}}
  2015 - 2015 {{IEEE Region}} 10 {{Conference}}}, pp.~1--6, Nov. 2015.

\bibitem{colomboCMOS253Ppmdeg2012}
D.~Colombo, F.~Werle, G.~Wirth, and S.~Bampi, ``A {{CMOS}} 25.3
  ppm{$^\circ$}/{{C}} bandgap voltage reference using self-cascode composite
  transistor,'' in {\em 2012 {{IEEE}} 3rd {{Latin American Symposium}} on
  {{Circuits}} and {{Systems}} ({{LASCAS}})}, pp.~1--4, Feb. 2012.

\bibitem{giustolisiLowvoltageLowpowerVoltage2003}
G.~Giustolisi, G.~Palumbo, M.~Criscione, and F.~Cutri, ``A low-voltage
  low-power voltage reference based on subthreshold {{MOSFETs}},'' {\em IEEE
  Journal of Solid-State Circuits}, vol.~38, pp.~151--154, Jan. 2003.

\bibitem{ueno300NW152009}
K.~Ueno, T.~Hirose, T.~Asai, and Y.~Amemiya, ``A 300 {{nW}}, 15
  ppm/{\textsuperscript{{\textbackslash}circ}}{{C}}, 20 ppm/{{V CMOS Voltage
  Reference Circuit Consisting}} of {{Subthreshold MOSFETs}},'' {\em IEEE
  Journal of Solid-State Circuits}, vol.~44, pp.~2047--2054, July 2009.

\bibitem{jiang65nmCMOSLow2018}
J.~Jiang, W.~Shu, and J.~S. Chang, ``A 65-nm {{CMOS Low Dropout Regulator
  Featuring}} {$>$}60-{{dB PSRR Over}} 10-{{MHz Frequency Range}} and 100-{{mA
  Load Current Range}},'' {\em IEEE Journal of Solid-State Circuits}, vol.~53,
  pp.~2331--2342, Aug. 2018.

\bibitem{el-nozahiHighPSRLow2010}
M.~{El-Nozahi}, A.~Amer, J.~Torres, K.~Entesari, and E.~{Sanchez-Sinencio},
  ``High {{PSR Low Drop-Out Regulator With Feed-Forward Ripple Cancellation
  Technique}},'' {\em IEEE Journal of Solid-State Circuits}, vol.~45,
  pp.~565--577, Mar. 2010.

\end{thebibliography}

\end{document}